\documentclass{PoS}

\title{A view of the sub-mJy populations, modelling and perspectives for 
future deep surveys}

\ShortTitle{A view of the sub-mJy populations}

\author{\speaker{Isabella Prandoni}, Paola Parma\\
        INAF - Istituto di Radioastronomia, Via Gobetti 101, Bologna (Italy)\\
        E-mail: \email{prandoni@ira.inaf.it}}
\author{Arturo Mignano\\
        INAF - Oss. Astronomico di Brera, Via Brera 28, Milano (Italy)}
\author{Hans R. de Ruiter\\
        INAF - Oss. Astronomico di Bologna, Via Ranzani 1, Bologna (Italy)}
\author{Loretta Gregorini \\
        Dip. Fisica, Univ. degli studi di Bologna, Via Irnerio 46, Bologna 
        (Italy)}
\author{Giampaolo Vettolani\\
         INAF, Viale del Parco Mellini 84, Roma (Italy)}
\author{Mark H. Wieringa and Ron D. Ekers\\
        CSIRO Australia Telescope Facility, PO Box 76, Epping NSW 2121 
        (Australia)}

\abstract{We use deep multi--colour (UBVRIJK) images mostly taken in 
the framework of the ESO \emph{Deep Public Survey} (DPS) to optically
identify and derive photometric redshifts for a complete sample of 131 radio 
sources with $S>0.4$ mJy, observed at both 1.4 
and 5 GHz as part of the ATESP radio survey. The availability of 
multi--wavelength radio and optical information is exploited to infer
the physical properties of the faint radio population. In particular we find 
that, considering both early-type galaxies and quasars as 
sources with an active nucleus, AGNs largely dominate our sample sub-mJy
sample (78\%). Further radio/optical analysis of such AGN component 
has revealed a somewhat unexpected class of flat/inverted--spectrum 
sources with low radio--to--optical ratios ($R<100$),
which are preferentially identified with early--type galaxies. Such
sources are quite compact ($d<10-30$ kpc), suggesting core-dominated radio 
emission triggered by low luminosity AGNs. This
intriguing class of objects deserves further analysis, and new higher 
resolution radio observations are currently under way. In parallel we are
developing radio source models, for both the AGN and the star-forming 
components of the sub--mJy radio pupulation. Here we discuss the first results.
}

\FullConference{From Planets to Dark Energy: the Modern Radio Universe\\
		 October 1-5 2007\\
		 The University of Manchester, UK}

\begin{document}

\section{Scientific Background}
One of the most debated issues about the sub-milliJy
(sub-mJy) radio sources responsible for the steepening of the 1.4~GHz
source counts (\cite{Condon1984}, \cite{Windhorst1990}) is the origin of 
their radio emission.  There
are strong theoretical and observational reasons, based also on
studies at other wavelengths, to believe that such sources are mainly
rapidly evolving star-forming galaxies, 
but there is also observational evidence that the contribution
of other classes of objects in the radio below 1 mJy is
important. Multi-wavelength studies of deep radio fields show that
star-forming galaxies dominate the microJy ($\mu$Jy) population (see
e.g. \cite{Richards1999}) while early-type galaxies and AGN
become dominant at flux 
densities $>0.1-0.2$ mJy (see e.g. \cite{Gruppioni1999}, 
\cite{Magliocchetti2000},
\cite{Prandoni2001b}, \cite{Afonso2006}, \cite{Mignano2007b}).
Large numbers of 
so-called radio-intermediate quasars have
been found at mJy levels (see e.g. \cite{Lacy2001}) and 
Jarvis \& Rawlings (\cite{Jarvis2004}) 
propose a scenario in which radio-quiet quasars strongly affect the
radio counts in the flux range $0.3-1$ mJy. However, the picture
outlined above needs further confirmation since most efforts so far have
been devoted to the study of $\mu$Jy samples, where star-forming
galaxies dominate. \\
Assessing whether the AGN-triggered component of the 
sub--mJy population is more related to efficiently accreting systems -
like radio-intermediate/quiet quasars - or to systems with very low accretion
rates - like e.g. FRI (\cite{Fanaroff1974}) 
radio galaxies - would have a relevant impact
on the study of the physical and evolutionary properties of low-power radio 
AGNs, and, more generally, would allow us a better 
understanding of the triggering mechanisms of AGN radio activity.

\begin{figure}[t]
\begin{center}
\resizebox{12cm}{!}{\includegraphics[]{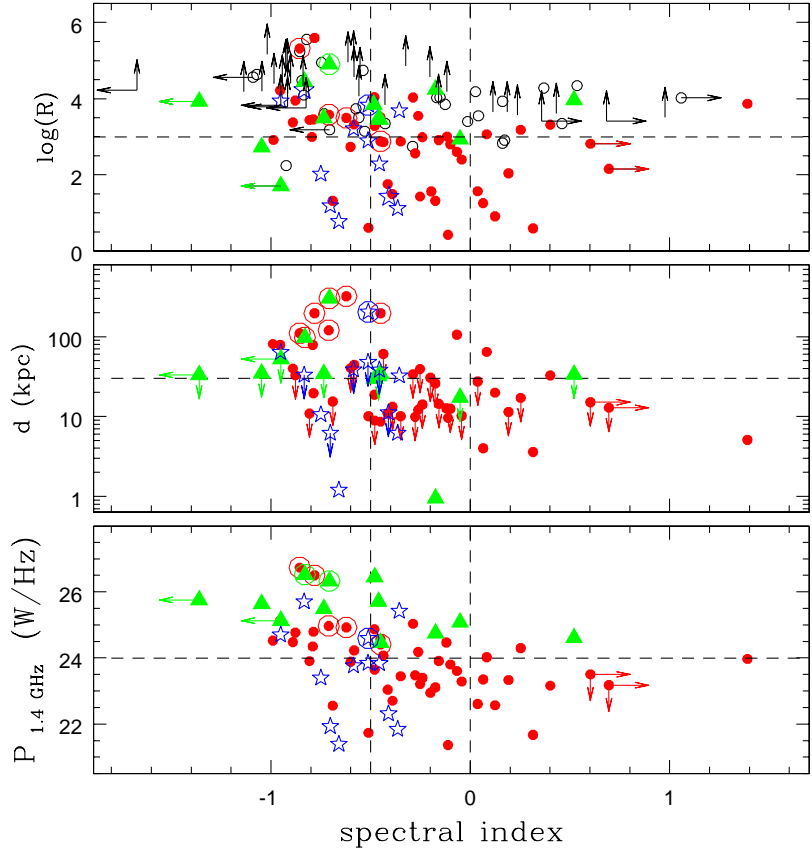}}
\caption{Radio--to--optical 
ratio ($R$, top panel), linear radio
source size ($d$ in kpc, middle panel) and 1.4 GHz radio power
(in W/Hz, bottom panel) against 1.4 -- 5 GHz spectral index 
for the ATESP-DEEP1 radio sources: 
red filled circles (ETS); blue stars 
(LTS/SB); green filled triangles (AGNs). Circled
symbols indicate sources with extended and/or two-component radio morphology,
typical of classical radio galaxies. 
Black open circles are for identified sources which do not have a 
redshift/type determination. Arrows indicate upper/lower limits. 
Vertical dashed lines indicate the $\alpha =-0.5$ and the $\alpha =0$ 
values, above which source spectra are defined respectively 
as flat and inverted. Horizontal dashed lines in the three panels indicate 
respectively (from top to bottom)
values of $R=1000$, $d=30$ kpc and $P_{\rm 1.4 GHz}=10^{24}$ W/Hz.
\label{fig:si_plots}}
\end{center}
\end{figure}

\section{Observational Results: the ATESP--DEEP1 Sample}
One especially suited radio sample to study the phenomenon of low-luminosity 
nuclear activity, possibly related to 
low efficiency accretion processes and/or radio-intermediate/quiet QSOs, 
is the so--called ATESP--DEEP1 sample, which consists of 131 radio sources 
with $S_{\rm lim}\sim 0.4$ mJy,
covering a one square degree area and imaged at both 5 and 1.4~GHz 
in the framework of the Australia Telescope ESO Slice Project (ATESP) 
survey (see \cite{Prandoni2000a}, \cite{Prandoni2000b}, \cite{Prandoni2006}). 
It is worth to note that at 
the flux densities probed here (mainly $0.5 - 1$ mJy) star--forming galaxies 
should start to contribute but should not yet be the dominant population and, 
in addition, this sub--mJy sample is one of the few with multi--frequency 
radio information.
Deep ($I<24$) UBVRIJK multi-color imaging is also available from the 
ESO Deep Public Survey (DPS, see \cite{Mignano2007a}, 
\cite{Olsen2006}), allowing to identify and get a redshift 
determination for about 80\% of the radio sources (see \cite{Mignano2007b}).
We confirm that at the sub-mJy level the large 
majority of sources are associated with objects that have
early--type (64\%) and AGNs (14\%) spectra. 
Although earlier work (based on shallower optical follow-up) 
revealed the presence of a conspicuous component of 
late--type and star-burst objects, such objects appear to be important only 
at bright magnitudes ($I<19$), and are rare at fainter magnitudes. \\
From an overall comparison of the radio spectral index with other 
radio and optical properties of the ATESP--DEEP1 sample,
we find that most sources with flat radio spectra have high radio-to-optical 
ratios, as expected for classical radio galaxies and quasars, while 
star-forming galaxies are associated to steep-spectrum radio sources. 
However the multi--frequency/multi--band analysis has 
revealed a somewhat unexpected class of objects with
flat and/or inverted spectrum {\it and} low radio-to-optical 
ratios ($10<R<1000$, see Fig. 1, top panel). Such sources are compact, with 
linear sizes $d< 10-30$ kpc (see Fig. 1, middle panel), and associated to 
optically inactive early-type galaxies (see \cite{Mignano2007b}). 
The distribution in redshift of these galaxies extends up 
to z$\sim2$, showing a significant peak at z=0.5, indicating that such 
sources may undergo significant evolution. \\
Their radio luminosities 
($P_{1.4 \rm{GHz}}\sim 10^{22-24}$ WHz$^{-1}$, see Fig. 1, bottom panel) 
and the absence of emission lines in their optical spectra may suggest that 
these objects belong to the class of FRI
radio galaxies. FRI radio galaxies, however, are characterized on average 
by steeper radio spectra and larger linear sizes. The 
compactness of the sources, together with the
flat/inverted spectra, suggests core emission with strong synchrotron 
or free-free self-absorption. Such sources may therefore either 
represent a specific
sub-class of core-dominated FRI radio galaxies, or may be associated to 
specific phases in the life of a radio source, or also may be similar to the 
low power compact radio sources discussed by Giroletti et al. 
(\cite{Giroletti2005}). 
 
\begin{figure*}[t]

\vspace{-4cm}
\begin{center}
\resizebox{12cm}{!}{\includegraphics[]{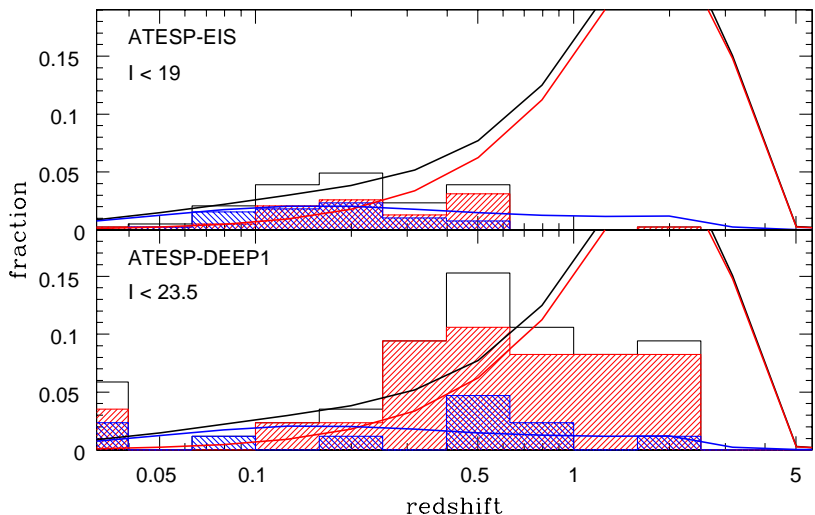}}
\caption{Redshift distribution for the shallower ($I<19$) ATESP-EIS sample
(top panel) and for the deeper ($I<24$) ATESP-DEEP1 sample (bottom panel).
The observed redshift distributions (black histograms) are compared to the 
modeled ones (black lines). Red lines and histograms indicate the AGN 
fraction (steep + flat components); Blue lines and histograms indicate the 
star-forming galaxy fraction (evolving + non-evolving components).
\label{fig:atesp-models}}
\end{center}
\end{figure*}

\section{Comparison with Models}
In order to probe the evolution of low-luminosity radio AGNs we have compared 
our observational results with standard models developed for the faint radio 
population. We have modeled the AGN component following Dunlop \& Peacock 
(\cite{Dunlop1990}) in two sub-components: steep and flat AGNs.
The star-forming galaxies have been modeled starting from the local radio 
luminosity function of Sadler et al. (\cite{Sadler2002}) in two sub-components:
non-evolving normal spirals and evolving starburst galaxies 
($L\sim (1+z)^3$).\\
In Fig.~2 we compare the models to the source redshift distribution of 
the shallower ($I<19$) ATESP-EIS 3~sq.~degr. sample (studied in 
\cite{Prandoni2001b}, top panel) and of the deeper ($I<24$) ATESP--DEEP1 
sample (bottom panel). The Figure clearly shows that only going
to deeper optical magnitudes we can really probe the models developed 
for the faint radio population and that standard modeling can reproduce with 
good accuracy the redshift distribution of the ATESP--DEEP1 sample, for both
the AGN and star-forming galaxy components. 

\section{Conclusions and Future Perspectives}

In this paper we have discussed the nature of the sub-mJy radio 
population, through the radio/optical analysis of the ATESP--DEEP1 sample.
The high percentage of identifications ($\sim 78\%$) allowed us to make 
a direct comparison between observations and models, in order to assess 
the nature and the evolutionary properties of the sub-mJy population, with 
particular respect to the AGN component. 
Our radio/optical analysis has revealed a class of flat/inverted spectrum 
radio sources associated to optically inactive early-type galaxies, which 
are most plausibly triggered by low-accretion/radiative efficiency AGNs. 
In order to further investigating this class of objects higher resolution 
multi-frequency radio observations are on-going.\\
We find a general agreement between observations and standard pure luminosity
evolution radio source modeling.  
It is clear however that, to provide more quantative constraints to the models,
we need larger deep radio samples together with complete optical follow--up.
In particular deeper radio surveys are necessary to probe a possible 
radio-quiet AGN component, which does not show up at the flux densities probed 
by our sample ($S>0.4$ mJy). \\
A clear understanding of the nature and evolutionary properties of the various
components of the sub--mJy population is a very important issue for providing
reliable modeling of the nanoJy radio sky will be eventually probed by the SKA,
and will be importat matter of study for the new upcoming radio facilities
(e.g. LOFAR, EVLA, eMERLIN, etc.).

\end{document}